\documentclass[11pt,letterpaper]{article}
\usepackage[english]{babel}
\usepackage[utf8]{inputenc}
\usepackage[T1]{fontenc}
\usepackage[colorlinks = true, citecolor=blue, urlcolor=blue]{hyperref}
\usepackage{enumitem}
\usepackage{amsmath,amsfonts,amssymb,bm}
\usepackage{graphicx}
\usepackage{tabularx}
\usepackage{booktabs}
\usepackage{dcolumn}
\usepackage{adjustbox}
\usepackage{caption}
\usepackage{xcolor}
\usepackage{authblk}
\usepackage{fullpage}
\usepackage{orcidlink}
\usepackage{enumitem}
\usepackage{xurl}

\DeclareMathOperator*{\argmin}{\arg\!\min}
\makeatletter
\newcolumntype{B}[3]{>{\boldmath\DC@{#1}{#2}{#3}}c<{\DC@end}}
\makeatother
\newcommand{\mybold}[1]{\multicolumn{1}{B{.}{.}{2.3}}{#1}}

\title{Determinants of renewable energy consumption in Madagascar:\authorcr
	Evidence from feature selection algorithms}

\author{Franck M.~Ramaharo\orcidlink{0000-0003-0340-170X}\thanks{Corresponding author: \href{mailto:franck.ramaharo@univ-antananarivo.mg}{\tt franck.ramaharo@univ-antananarivo.mg}}\, and Fitiavana M.~Randriamifidy\orcidlink{0009-0003-3644-4455}}

\affil{\small Département de Mathématiques et Informatique \authorcr
	Université d'Antananarivo\authorcr
	Antananarivo 101, Madagascar
}

\date{\small\today}

\newcommand{\REC}{\mathit{REC}}
\newcommand{\CO}{\mathit{CO}2}
\newcommand{\DINV}{\mathit{DINV}}
\newcommand{\EG}{\mathit{EG}}
\newcommand{\EXR}{\mathit{EXR}}
\newcommand{\FDEV}{\mathit{FDEV}}
\newcommand{\FDI}{\mathit{FDI}}
\newcommand{\INC}{\mathit{INC}}
\newcommand{\IND}{\mathit{IND}}
\newcommand{\INFL}{\mathit{INFL}}
\newcommand{\TOUR}{\mathit{TOUR}}
\newcommand{\TR}{\mathit{TR}}
\newcommand{\URB}{\mathit{URB}}

\begin{document}
	\maketitle
	\begin{abstract}
		The aim of this note is to identify the factors influencing renewable energy consumption in Madagascar. We tested 12 features covering macroeconomic, financial, social, and environmental aspects, including economic growth, domestic investment, foreign direct investment, financial development, industrial development, inflation, income distribution, trade openness, exchange rate, tourism development, environmental quality, and urbanization. To assess their significance, we assumed a linear relationship between renewable energy consumption and these features over the 1990–2021 period. Next, we applied different machine learning feature selection algorithms classified as filter-based (relative importance for linear regression, correlation method), embedded (LASSO), and wrapper-based (best subset regression, stepwise regression, recursive feature elimination, iterative predictor weighting partial least squares, Boruta, simulated annealing, and genetic algorithms) methods. Our analysis revealed that the five most influential drivers stem from macroeconomic aspects. We found that domestic investment, foreign direct investment, and inflation positively contribute to the adoption of renewable energy sources. On the other hand, industrial development and trade openness negatively affect renewable energy consumption in Madagascar.
		
		\bigskip
		
	\end{abstract}
	
	\section{Introduction}
	
	The Malagasy government has adopted Madagascar’s \textit{New Energy Policy} (NEP) for 2015--2030, and aims to increase the household access rate to modern and affordable electricity from 15\% to 70\% by 2030 while increasing the share of renewable energies to 85\% in the energy mix \cite{MEH2015}. This energy policy recognizes that rural electrification plays a crucial role in achieving the first target, either by extending the power grid, building mini-grids, or providing individual off-grid solutions \cite{ARCEB2022, Knowledge4Policy, MEH2015a, MEH2021, seforall2019}. The second target of the NEP takes into account the high potential for renewable energy sources on the island (hydro, solar, wind, and biomass) and aims to fully harness these resources to supply clean and sustainable energy. Various projection scenarios and investment plans have been proposed to assess the feasibility of such a project \cite{DutoitMcLachlan, MEH2015a, MinWEH2018, Praeneetal2017, Praeneetal2021, Randrianariveloetal2022, RasamoelinaPraene2018}.

	This note represents a modest but meaningful step toward identifying the primary factors influencing the demand for renewable energy in Madagascar. These factors encompass macroeconomic, financial, social, and environmental aspects as addressed by the NEP. For this purpose, we initially identified 12 key drivers of renewable energy use that are commonly found in both theoretical and empirical analyses \cite{Ansarietal2021, Azam2016, BamatiandRaoofi2020, Banoetal2021, BenJebli2019, Foye2023, Gozgoretal2020, GrossetandNguyen-Van, Iqbaletal2023, Islametal2022, Maliketal2014, MohamedYusoffetal2023, MurshedAlam2021, Samouretal2022, Sunetal2023}, which include economic growth ($\EG$), domestic investment ($\DINV$), foreign direct investment ($\FDI$), financial development ($\FDEV$), industrial development ($\IND$), inflation ($\INFL$), income distribution ($\INC$), trade openness ($\TR$), exchange rate ($\EXR$), tourism development ($\TOUR$), environmental quality ($\CO$) and urbanization ($\URB$). An empirical model for investigating how these factors affect renewable energy demand in Madagascar can be defined as follows:
	\begin{equation}\label{eq:lin_rel}
		\REC = f(\CO, \DINV, \EG, \EXR, \FDEV, \FDI, \INC, \IND, \INFL, \TOUR, \TR, \URB),
	\end{equation}
	where $\REC$ represents renewable energy consumption. Our strategy begins by assuming that the model \eqref{eq:lin_rel} follows a linear form. Next, we implement different popular machine learning algorithms, namely feature selection algorithms, to determine which of these factors exert the greatest impact on renewable energy consumption in Madagascar. The outcome of this process will yield insights into which specific factors the government should prioritize to effectively stimulate the demand for renewable energy.

	\section{Energy transition challenges and opportunities in Madagascar: a focus on macroeconomic, financial, social, and environmental factors}
	
	\subsection{Macroeconomic factors}
	
	In general, developing countries have a strong dependence on imported fossil fuels, so their energy supply is vulnerable to fluctuations in global oil and gas prices. When fossil fuel prices rise, it can lead to energy supply disruptions and increased costs for consumers \cite{SurroopRaghoo2018}. In Madagascar, inflation is mainly driven by energy and food prices \cite{AfDB2023}, which contribute 53\% and 20\% (in fact, the two largest shares), respectively, to the consumer price index \cite{DSCM2019,INSTATIPC2022a,INSTATCN2022c}.  In response, households and small and medium-sized enterprises  (SMEs) seek more stable and cost-effective alternatives, including renewable energy. The exchange rate also plays a crucial role because it can increase the cost of importing renewable energy equipment and technology, making them more expensive for consumers and businesses. However, exchange rate depreciation can also increase economic activity, which in turn can lead to higher energy consumption and renewable energy penetration \cite{Foye2023, Shahetal2022}.

	The dependency of Madagascar's economy on imported fossil fuels can be observed through its industrial development and foreign trade structure.  According to the estimation of the National Institute of Statistics of Madagascar (INSTAT), the share of the industrial sector in Madagascar's gross domestic product in 2021 was 14.7\%. Industry value added has recorded an annual average percent growth rate of 3.7\%  from 2007 to 2021 and is largely dominated by mining and textiles \cite{INSTATCN2022b} (it should be noted that the mining and textile industries are responsible for around 69\% of the total volume of exports of goods \cite{Douane2023}). During the same period, the industrial sector accounted for approximately 65.3\% of the total medium voltage electricity consumption \cite{JIRAMA2023, RamaharoRazanajatovo2023}. JIRAMA (Jiro sy Rano Malagasy), Madagascar’s national electric utility and water services provider, generates electricity through thermal power plants fueled by diesel and heavy fuel oil, representing 20\% of the country’s total oil consumption.\cite{OMH2021, OMH2023}. The Ambatovy Project and QMM (QIT Madagascar Minerals), which represent not only the two major industrial mining operations in Madagascar but also the largest source of foreign direct investment for the country \cite{Maalejetal2018}, contributed approximately 9.5\% of the total oil consumption. About 22.1\% of the total volume of imported goods is dedicated to meeting these energy demands \cite{Douane2023}. These observations suggest that Madagascar's economic development relies heavily on fossil energy. However, investments in cleaner energy alternatives are also emerging, especially in the textile sector \cite{Cieltextile2022, EDBM2021}. The transition to renewable energy sources clearly shows the commitment of the industrial sector to promote economic growth in Madagascar, while addressing key issues related to access to energy and environmental protection. These investments may originate from both domestic and foreign sources.

	Domestic investment is manifested in the financing of key infrastructure projects such as hydroelectric dams, solar power plants, and wind turbines, thus increasing the country's capacity for renewable energy production. In addition, it facilitates research and development activities that should lead to technological innovations that enhance the efficiency and cost-effectiveness of renewable energy sources. \cite{MESUPRES2015}. Madagascar actively promotes private investment in clean and sustainable energy, offering generous tax incentives such as exemptions from VAT and import duties for solar panels, wind turbines, and batteries \cite{PRC-ELEC2019}. Moreover, Madagascar's government recognizes that public-private partnerships are crucial for advancing rural electrification projects and creating a favorable legal framework for decentralized renewable energy. For example, ``Volobe'' and ``Sahofika'' are two of the most important hydropower projects, which have a capacity of 120 MW and 192 MW of renewable electricity, respectively \cite{DCRPPresidence2021, DCRPPresidence2023}.

	Initiating a shift toward renewable energy is a complex endeavor, and developing countries often rely on foreign direct investments (FDIs) to kick-start their renewable energy sectors \cite{Benli2021, Darwinetal2022, Polat2018}. In fact, FDI can bring capital and technology transfer, and in the case of renewable energy, it can introduce advanced technologies and expertise, making it easier to exploit renewable resources efficiently. For example, Rio Tinto, the world's leading mining and materials company, which owns 80\% of QMM, has taken significant steps to reduce its carbon footprint in Madagascar by investing in renewable energy sources. As part of a commitment of USD 7.5 billion to reduce carbon emissions by 2023, the mining giant has launched the construction of solar and wind power plants on the island. The solar power plant, together with an energy storage system equipped with lithium-ion batteries, will enable Rio Tinto to cover up to 60\% of its annual electricity consumption \cite{Tossou2021}.

	\subsection{Financial factors}
	Financial development can positively impact renewable energy consumption through increased investment, risk mitigation, technology support, wider access to renewable solutions, and alignment with government policies. National financial institutions with insurance and state guarantees should participate in financing renewable energy, fostering a conducive environment for cleaner energy adoption, and addressing energy access challenges. Two notable examples of financial initiatives in this domain are the Off-Grid Market Development Fund (OMDF) and the Sustainable Use of Natural Resources and Energy Finance (SUNREF). These programs focus on providing financial solutions and support to promote sustainable and environmentally friendly projects, with a particular emphasis on renewable energy, energy efficiency, and environmental initiatives \cite{PERER2020}. OMDF is dedicated to expanding access to electricity through off-grid solar energy solutions. Supported by Madagascar's government and in partnership with the World Bank, this program is managed by Bamboo Capital Partners. It offers a results-based financing grant program that serves as an incentive for solar companies to expand their operations and achieve specific distribution targets. Furthermore, OMDF extends credit solutions to assist distributors and financial institutions engaged in the off-grid solar sector, facilitating working capital management and inventory control. The primary goal of OMDF is to increase electricity access, particularly in underserved areas, by delivering solar products such as solar lamps and solar home systems to households and small businesses, to reach at least 300,000 households and SMEs by 2024 \cite{OMDF2020}. SUNREF is a financial and technical assistance program initiated and funded by the French Development Agency (AFD). The primary mission of this program is to promote green financing, reduce carbon emissions, and encourage sustainable practices, ultimately contributing to economic and environmental sustainability in the countries where it operates \cite{SUNREF}. SUNREF provides partner banks with a favorable line of credit to finance environmentally friendly investments, aids businesses and banks in identifying and implementing sustainable projects, and is committed to improving energy security, enhancing business access to bank financing, and reducing obstacles to the implementation of renewable energy, energy efficiency, and environmental performance initiatives. SUNREF Madagascar plays a pivotal role in advancing sustainable practices and projects while contributing to economic growth \cite{MalagasyNews2021}.

	\subsection{Social and environmental factors}
	Income distribution plays a crucial role in adopting energy-related technologies and practices. Low-income households and rural people who do not have access to electricity rely on firewood, charcoal, oil, and batteries for their energy needs. In Madagascar, firewood and charcoal are the main sources of cooking energy. Access to electricity through decentralized electrification is limited to just 6.5\% of the rural population \cite{Lanetal2019}. For households lacking electricity access, kerosene lamps and candles are commonly used for lighting. The majority of these households allocate a substantial portion of their budgets to purchasing kerosene. Moreover, the existing technological solutions offered in the market are typically beyond the financial reach of a significant portion of the population. As a result, low-income consumers are unlikely to transition to a solution that lacks financial viability \cite{NaidooandLoot2020}. This situation often leads households to choose cost-effective rather than energy-efficient options, potentially driving up the demand for forest resources. Therefore, a significant portion of greenhouse gas emissions in Madagascar are due to anthropogenic activities, which include deforestation, slash-and-burn agriculture, charcoal production, and land use change \cite{Clark2012,MEEF2018}. Conversely, consumers with higher incomes are more inclined to transition to cleaner energy sources and technologies. This propensity is often driven by their greater financial capacity to invest in renewable energy systems and energy-efficient appliances. Additionally, higher-income consumers are often more environmentally conscious and willing to pay a premium for clean energy options \cite{seforall2019}.

	In addition to income distribution, spatial distribution also determines energy use. Urbanization in Madagascar has led to a significant increase in electricity demand; however, as of 2021, the country still faces a severe electricity access problem, with only 72.6\% of the urban population having access to this essential service \cite{WB_ACCESS}. Load shedding, characterized by prolonged power cuts, has become a common occurrence, affecting both households and businesses, particularly in large cities. To address this issue, authorities are turning to renewable energy solutions \cite{DCRPPresidence2021,DCRPPresidence2023}. Furthermore, rapid urbanization in developing countries leads to elevated CO2 emissions in urban areas \cite{AdusahPoku2016,Effiong2016}. In the case of Madagascar, this circumstance is caused by transportation demands that drive up the consumption of fossil fuels. Indeed, according to the Malagasy Office of Hydrocarbons, transportation-related fuel consumption constituted an average of 50.1\% of total oil consumption in the 2017-2021 period \cite{INSTATCN2022c, OMH2023, RamaharoRazanajatovo2023}. This situation highlights the urgency of implementing sustainable transportation solutions and transitioning to cleaner energy sources .

	In the field of ecotourism and sustainability, the use of renewable energy sources is also crucial \cite{Banoetal2021, BenJebli2019}. The tourism sector in developing countries is heavily dependent on fossil fuels and has harmful environmental consequences. As a result, energy use needs to be integrated into the debate on sustainable tourism development \cite{Gossling2010}. This form of responsible travel promotes conservation efforts, strengthens the economy and provides Malagasy with job opportunities. Renewable energy, in particular solar and wind energy, plays a key role in reducing the environmental impact of tourism activities, reducing the environmental impact of tourism on fragile ecosystems and protecting the natural environment of the country. In addition, the use of renewable energies contributes to the economic well-being of local communities and provides them with sustainable and environmentally friendly energy sources \cite{Casseetal2022}. As a result, a number of hotels in Madagascar are encouraged to implement sustainable practices, such as the exclusive use of renewable energy sources, which contributes to the growing trend towards eco-tourism \cite{HafalianavalonaExpress2017, SOLIDIS2017}.

	\section{Data and empirical methodology}

	We set an indicator for each factor defined in \eqref{eq:lin_rel}, see Table \ref{tab:variable_description}. Our data range from 1990 to 2021 and are mainly obtained from the National Institute of Statistics (INSTAT), the Central Bank of Madagascar (BFM), the World Development Indicators (WDI), and Our World In Data (OWID).

	\begin{table*}[htbp]
		\centering
		\caption{Variables description.}
		\label{tab:variable_description}
		\begin{tabular}{@{}p{0.08\linewidth}p{0.675\linewidth}p{0.175\linewidth}@{}}
			\toprule
			Variables & Indicators description and measurement unit & Source\\
			\midrule
			$\REC$ & Renewable energy consumption  (\% of total final energy consumption) & WDI \cite{WDI_REC} \\
			$\CO$ & CO2 emission per capita (annual percent change)  &  OWID \cite{OWID2023,Ritchieetal2020} 	\\
			$\DINV$ & Gross fixed capital formation (\% of GDP) & INSTAT \cite{INSTATCN2022b,INSTATCN2022c}\\
			$\EG$& Real gross domestic product  (annual percent change) & INSTAT \cite{INSTATCN2022b,INSTATCN2022c}\\
			$\EXR$ & Period average exchange rate USD/MGA (annual percent change) & BFM \cite{BFM2023b, Ramaharoetal2023}\\
			$\FDEV$ & Domestic credit to private sector (\% of GDP) & BFM \cite{BFM2023a, BFM2023c} \\
			$\FDI$ & Foreign direct investment, net inflows (\% of GDP) & BFM \cite{BFM2023a, BFM2023c} \\
			$\INC$ & Gross disposable private income (\% of GDP) & INSTAT 	\cite{INSTATCN2022b}\\
			$\IND$ & Industry value added  (annual percent change) & INSTAT \cite{INSTATCN2022b,INSTATCN2022c} \\
			$\INFL$& Period average consumer price index (annual percent change) & INSTAT \cite{INSTATIPC2022a,INSTATCN2022c}\\
			$\TR$& Sum of exports and imports of goods and non-factor services (\% of GDP) & BFM \cite{BFM2023a, BFM2023c}\\
			$\TOUR$ & Number of tourist arrivals (annual percent change) & INSTAT \cite{INSTATCN2022c,Ramaharo2023}\\		
			$\URB$ & Urban population (annual percent change) &  WDI \cite{WDI_URB} \\
			\bottomrule
		\end{tabular}
	\end{table*}

	In this study, all variables were centered and scaled.  The statistical representation of \eqref{eq:lin_rel} then takes the following linear functional form:
	\begin{equation}\label{eq:linreg}
		\begin{split}
			\REC_t &= \beta_1\CO_t + \beta_2\DINV_t + \beta_3\EG_t + \beta_4\EXR_t + \beta_5\FDEV_t + \beta_6\FDI_t\\
			&+ \beta_7\INC_t + \beta_8\IND_t + \beta_{9}\INFL_t + \beta_{10}\TOUR_t + \beta_{11}\TR_t +\beta_{12}\URB_t + \varepsilon_t,
		\end{split}
	\end{equation}
	where $\beta_i,\, i= 1,\ldots, 12$ are unknown parameters of the regressors to be estimated, and $\varepsilon_t$ is the error term of the regression.

	Initially, we applied both multiple linear regression and principal component regression to estimate the model, incorporating all available features. We then run different algorithms to select the optimal features and re-estimate the model with the selected features.  Feature selection algorithms can be classified as filter-based (relative importance for linear regression, correlation method), embedded (LASSO), and wrapper-based (best subset regression, stepwise regression, recursive feature elimination, iterative predictor weighting partial least squares, Boruta, simulated annealing, and genetic algorithms) methods. Except for the LASSO model, each model is estimated using ordinary least squares (OLS). We use the \texttt{R} programming language, which already includes packages that implement these algorithms. Because of the small size of our sample, the entire dataset will serve as the training set for our models. When dealing with a model that has hyperparameters that need to be set, our preferred method is the k-fold cross-validation, which helps us evaluate the model's performance under different hyperparameter configurations. The dataset, along with the \texttt{R} script, is available for download on the Open Science Framework platform \cite{Ramaharo2023data}.

	\section{A feature selection problem}
	\subsection{Fitting a full model}
	\subsubsection{Multiple linear regression}
	We build the first model by estimating \eqref{eq:linreg} with ordinary least squares.
	\newcounter{mycounter}
	\setcounter{mycounter}{1}
	\begin{gather}\label{eq:regfull}
		\begin{split}
			\textbf{Mod\themycounter}:\quad \widehat{\REC_t} &= 
			-\underset{(0.13561)}{0.119419}\,{\CO_t}
			+\underset{(0.14309)^{**}}{0.374359}\,{\DINV_t}
			+\underset{(0.27362)}{0.0478255}\,{\EG_t}
			-\underset{(0.15094)}{0.0303502}\,{\EXR_t}\\
			&+\underset{(0.16190)^{**}}{0.352481}\,{\FDEV_t}
			+\underset{(0.20506)^{**}}{0.553816}\,{\FDI_t}
			-\underset{(0.11670)}{0.101220}\,{\INC_t}
			-\underset{(0.18947)^{**}}{0.307435}\,{\IND_t}\\
			&+\underset{(0.12920)^{*}}{0.269445}\,{\INFL_t}
			+\underset{(0.21919)}{0.262536}\,{\TOUR_t}
			-\underset{(0.19683)^{***}}{0.618646}\,{\TR_t}
			+\underset{(0.17221)}{0.0133764}\,{\URB_t}
		\end{split}
		\notag \\
		T = 32 \quad \bar{R}^2 = 0.7063 \quad F(12,19) = 7.2127^{***} \quad \hat{\sigma} = 0.54194 \quad DW = 1.411028 \notag \\
		\centerline{(standard errors in parentheses; $^{*}p<0.1$, $^{**}p<0.05$, $^{***}p<0.01$).} \notag
	\end{gather}

	\subsubsection{Principal component regression}

	Principal component regression (PCR) is a machine learning algorithm that combines dimensionality reduction using principal component analysis (PCA) with linear regression. First, PCA is applied to the feature space, transforming the original features into a set of orthogonal ``principal components'' that capture the most significant variance in the data while eliminating multicollinearity. Then, it performs linear regression on these principal components to build a predictive model. The linear combination property of these principal components ensures that the obtained predictive model can be interpreted in terms of the original features \cite{RamaharoetRajaonarison2023}.

	We applied k-fold cross-validation to determine the optimal number of principal components $\ell$ to retain in the PCR model. The value $\ell = 6$ corresponds to the minimum root mean square error of cross-validation (RMSECV). This value of $\ell$ accounts for 86.30\% of the total variability. 
	
	\begin{gather}\label{eq:regpcr}
		\begin{split}
			\textbf{Mod\stepcounter{mycounter}\themycounter}:\quad \widehat{\REC_t} &= 
			-0.110714\, \CO_t
			+0.325094\, \DINV_t
			-0.158226\, \EG_t
			-0.013973\, \EXR_t\\
			&+0.013010\, \FDEV_t
			+0.114446\, \FDI_t
			+0.037417\, \INC_t
			-0.132716\, \IND_t\\
			&+0.255160\, \INFL_t
			+0.056763\, \TOUR_t
			-0.243690\, \TR_t
			+0.411719\, \URB_t.
		\end{split}
		\notag
	\end{gather}

	\subsection{Correlation method}
	This method simply computes the Pearson correlation matrix, which allows us to filter out redundant variables. We set the pair-wise absolute correlation cutoff to $ 0.75 $ such that if two variables have a higher correlation than the cutoff, then the variable with the largest mean absolute correlation is removed.
	
	\begin{table}[htbp]
		\centering
		\caption{Pearson correlation matrix.}
		\begin{adjustbox}{width=\linewidth,center}
			\begin{tabular}{@{}lD{.}{.}{2.2}D{.}{.}{2.2}D{.}{.}{2.2}D{.}{.}{2.2}D{.}{.}{2.2}D{.}{.}{2.2}D{.}{.}{2.2}D{.}{.}{2.2}D{.}{.}{2.2}D{.}{.}{2.2}D{.}{.}{2.2}D{.}{.}{2.2}c@{}}
				\toprule
				& \multicolumn{1}{c}{$\REC$} & \multicolumn{1}{c}{$\CO$} & \multicolumn{1}{c}{$\DINV$} & \multicolumn{1}{c}{$\EG$} & \multicolumn{1}{c}{$\EXR$} & \multicolumn{1}{c}{$\FDEV$} & \multicolumn{1}{c}{$\FDI$} & \multicolumn{1}{c}{$\INC$} & \multicolumn{1}{c}{$\IND$} & \multicolumn{1}{c}{$\INFL$} & \multicolumn{1}{c}{$\TOUR$} & \multicolumn{1}{c}{$\TR$} & $\URB$\\
				\midrule
				$\REC$& 1&       &       &       &       &       &       &       &       &       &       &       &  \\
				$\CO$& -0.344 & 1&       &       &       &       &       &       &       &       &       &       &  \\
				$\DINV$& 0.502 & -0.200 & 1&       &       &       &       &       &       &       &       &       &  \\
				$\EG$& -0.395 & 0.460 & 0.168 & 1&       &       &       &       &       &       &       &       &  \\
				$\EXR$& -0.010 & 0.142 & -0.042 & -0.130 & 1&       &       &       &       &       &       &       &  \\
				$\FDEV$& 0.107 & 0.000 & -0.463 & -0.272 & 0.076 & 1&       &       &       &       &       &       &  \\
				$\FDI$& 0.153 & 0.010 & 0.274 & 0.035 & -0.275 & -0.087 & 1&       &       &       &       &       &  \\
				$\INC$& -0.112 & 0.150 & -0.272 & -0.169 & 0.055 & 0.164 & -0.092 & 1&       &       &       &       &  \\
				$\IND$& -0.449 & 0.496 & -0.003 & 0.818 & -0.046 & -0.188 & 0.094 & -0.036 & 1&       &       &       &  \\
				$\INFL$& 0.369 & -0.135 & 0.181 & -0.184 & 0.549 & -0.056 & -0.234 & 0.033 & -0.133 & 1&       &       &  \\
				$\TOUR$& -0.124 & 0.439 & 0.270 & 0.745 & 0.092 & -0.412 & -0.145 & -0.111 & 0.655 & 0.076 & 1&       &  \\
				$\TR$& -0.428 & 0.135 & 0.071 & 0.468 & -0.012 & -0.112 & 0.591 & -0.200 & 0.398 & -0.215 & 0.119 & 1&  \\
				$\URB$& 0.490 & -0.082 & 0.171 & -0.081 & -0.270 & 0.326 & 0.359 & -0.229 & -0.017 & -0.040 & 0.042 & -0.090 & 1\\
				\bottomrule
			\end{tabular}%
		\end{adjustbox}
		\label{tab:tab_cor}%
	\end{table}%
	
	As shown in Table \ref{tab:tab_cor}, $\EG$ and $\IND$ are the only pairs whose correlation coefficient, equal to $0.818$, exceeds the cut-off value. The mean absolute correlation of $EG$ is $0.321$, whereas that of $\IND$ is $0.203$. Therefore, $\EG$ is removed.  
	\begin{gather}\label{eq:regcor}
		\begin{split}
			\textbf{Mod\stepcounter{mycounter}\themycounter}:\quad \widehat{\REC_t} &= 
			-\underset{(0.12855)}{0.113828}\,{\CO_t}
			+\underset{(0.13072)^{***}}{0.383126}\,{\DINV_t}
			-\underset{(0.13159)}{0.0421862}\,{\EXR_t}
			+\underset{(0.15344)^{**}}{0.359177}\,{\FDEV_t}\\
			&+\underset{(0.19104)^{**}}{0.543191}\,{\FDI_t}
			-\underset{(0.11240)}{0.104448}\,{\INC_t}
			-\underset{(0.15242)^{*}}{0.288703}\,{\IND_t}
			+\underset{(0.12596)^{**}}{0.268723}\,{\INFL_t}\\
			&+\underset{(0.18623)}{0.281359}\,{\TOUR_t}
			-\underset{(0.16755)^{***}}{0.601845}\,{\TR_t}
			+\underset{(0.16439)}{0.00718235}\,{\URB_t}
		\end{split}
		\notag \\
		T = 32 \quad \bar{R}^2 = 0.7205 \quad F(11,20) = 8.2663^{***} \quad \hat{\sigma} = 0.52864 \quad DW = 1.446916 \notag \\
		\centerline{(standard errors in parentheses; $^{*}p<0.1$, $^{**}p<0.05$, $^{***}p<0.01$).} \notag
	\end{gather}

	\subsection{Least absolute shrinkage and selection operator}

	The least absolute shrinkage and selection operator (LASSO) is a technique that can select variables based on their importance through shrinkage to simplify linear regression models and prevent overfitting. It imposes a penalty to reduce the absolute value of the magnitude of the regression coefficients, meaning that some of the coefficients will exhibit values close to zero or set to zero if they are deemed to be less important \cite{Tibshirani1996}. The optimization problem for the LASSO is given by the Lagrangian form
	\begin{equation}
		\argmin\limits_{\beta =(\beta_1, \ldots,\beta_p)} \left\{\sum_{i=1}^{T}\left(y_i - \sum_{j=1}^{p}\beta_j x_{ij}\right)^2    + \lambda\sum_{j=1}^{p} |\beta_j|\right\},
	\end{equation}
	where $\lambda$ denotes the penalty or hyperparameter optimization that controls the amount of shrinkage.  Note that if $\lambda = 0$, then the problem is reduced to ordinary least squares.

	We use  k-fold  cross-validation to tune $\lambda$  and the resulting value for $\lambda$ is $0.07686471$.
	\begin{gather}\label{eq:reglasso}
		\begin{split}
			\textbf{Mod\stepcounter{mycounter}\themycounter}:\quad \widehat{\REC_t} &= 
			0.358126\, \DINV_t    
			+ 0.051462\, \FDEV_t
			+ 0.113563\, \FDI_t     
			-0.230040\, \IND_t\\
			&+ 0.175139\, \INFL_t   
			-0.282611\, \TR_t
			+ 0.270503\, \URB_t.
		\end{split}
		\notag
	\end{gather}

	\subsection{Best subset regression}

	This procedure is a model selection approach that consists of testing all possible combinations of the regressors and then selecting the best subset according to some statistical criteria, usually adjusted R-squared, Mallows' $C_p$, Akaike information criterion (AIC), and Bayesian information criterion (BIC) \cite{Zhang2016}.

	We performed k-fold cross-validation to select the best subset. This process involves systematically evaluating various subsets of input features and selecting the subset that offers the best model performance while accounting for cross-validation errors. This approach allows the identification of the subset of input features that maximizes the model’s predictive performance during cross-validation and helps prevent overfitting by assessing how well the model generalizes to unseen data.
	\begin{gather}\label{eq:regbsr}
		\begin{split}
			\textbf{Mod\stepcounter{mycounter}\themycounter}:\quad \widehat{\REC_t} &= 
			\underset{(0.11902)^{***}}{0.454109}\,{\DINV_t}
			+\underset{(0.11099)^{***}}{0.346965}\,{\FDEV_t}
			+\underset{(0.13191)^{***}}{0.519819}\,{\FDI_t}
			-\underset{(0.14303)^{**}}{0.320693}\,{\IND_t}\\
			&+\underset{(0.099191)^{**}}{0.243405}\,{\INFL_t}
			+\underset{(0.14942)}{0.231586}\,{\TOUR_t}
			-\underset{(0.12763)^{***}}{0.575813}\,{\TR_t}
		\end{split}
		\notag \\
		T = 32 \quad \bar{R}^2 = 0.7360 \quad F(7,24) = 13.349^{***} \quad \hat{\sigma} = 0.51377  \quad DW = 1.393969 \notag \\
		\centerline{(standard errors in parentheses; $^{*}p<0.1$, $^{**}p<0.05$, $^{***}p<0.01$).} \notag
	\end{gather}

	\subsection{Stepwise regression}
	Stepwise regression is a feature selection technique used in machine learning and statistical modeling and is characterized by its iterative approach to building a predictive model. It operates by systematically adding or removing input features from the model based on specific criteria, often optimizing goodness-of-fit metrics such as AIC or BIC \cite{Zhang2016}. In forward selection, the method starts with an empty model and adds features one by one, selecting those that contribute the most to improving the chosen criterion. Conversely, backward elimination begins with a model having all features and removes them one at a time, selecting those whose removal enhances the criterion the most.  
	
	We apply both forward and backward selection and use AIC for our analysis, which quantifies the trade-off between model complexity and goodness-of-fit.
	\begin{gather}\label{eq:regstep}
		\begin{split}
			\textbf{Mod\stepcounter{mycounter}\themycounter}:\quad \widehat{\REC_t} &= 
			-\underset{(0.11518)}{0.144917}\,{\CO_t}
			+\underset{(0.12328)^{***}}{0.407550}\,{\DINV_t}
			+\underset{(0.11010)^{***}}{0.359284}\,{\FDEV_t}
			+\underset{(0.13242)^{***}}{0.549217}\,{\FDI_t}\\
			&-\underset{(0.14379)^{*}}{0.287367}\,{\IND_t}
			+\underset{(0.098196)^{**}}{0.235756}\,{\INFL_t}
			+\underset{(0.15674)^{*}}{0.297807}\,{\TOUR_t}
			-\underset{(0.12675)^{***}}{0.591782}\,{\TR_t}
		\end{split}
		\notag \\
		T = 32 \quad \bar{R}^2 = 0.7423 \quad F(8,23) = 12.162^{***} \quad \hat{\sigma} = 0.50764 \quad DW = 1.309463 \notag \\
		\centerline{(standard errors in parentheses; $^{*}p<0.1$, $^{**}p<0.05$, $^{***}p<0.01$).} \notag
	\end{gather}

	\subsection{Recursive feature elimination}
	Recursive feature elimination (RFE) is a feature selection method that enhances the performance of a model by systematically identifying and selecting the most relevant features from a dataset while eliminating less important ones \cite{ChenandJeong2007,Guyonetal2002}. It is a recursive process that starts with all the features in the dataset and computes an importance score for each regressor. Then, the selection process iteratively prunes the least important features from the current set until it has achieved the optimum number of features to keep. RFE trains the model during each iteration and evaluates its performance using a cross-validation strategy.
	\begin{gather}\label{eq:regrfe}
		\begin{split}
			\textbf{Mod\stepcounter{mycounter}\themycounter}:\quad \widehat{\REC_t} &= 
			\underset{(0.12909)^{***}}{0.428158}\,{\DINV_t}
			+\underset{(0.20497)}{0.155942}\,{\EG_t}
			+\underset{(0.12615)^{*}}{0.224403}\,{\FDEV_t}
			+\underset{(0.16794)^{**}}{0.407696}\,{\FDI_t}\\
			&-\underset{(0.17675)^{*}}{0.313252}\,{\IND_t}
			+\underset{(0.10499)^{**}}{0.275180}\,{\INFL_t}
			-\underset{(0.16372)^{***}}{0.547871}\,{\TR_t}
			+\underset{(0.12892)}{0.165951}\,{\URB_t}
		\end{split}
		\notag \\
		T = 32 \quad \bar{R}^2 = 0.7267 \quad F(8,23) = 11.306^{***} \quad \hat{\sigma} = 0.52274 \quad DW = 1.495053 \notag \\
		\centerline{(standard errors in parentheses; $^{*}p<0.1$, $^{**}p<0.05$, $^{***}p<0.01$).} \notag
	\end{gather}

	\subsection{Partial least squares regression}

	Partial least squares (PLS) regression consists of extracting a set of latent factors that explain as much of the covariance as possible between the response and the regressors, then performing least squares regression on these factors instead of on the original data \cite{GeladiKowalski1986}. There are three categories for the PLS-based feature selection method: filter, wrapper, and embedded methods \cite{Mehmoodetal2012,Mehmoodetal2022}.

	We used iterative predictor weighting partial least squares (IPW-PLS), which is an iterative elimination procedure in which a measure of regressor importance is computed after fitting a PLS regression model. It offers practical benefits in regression problems by eliminating useless predictors and producing acceptable regression models with a reduced number of relevant regressors \cite{Forina1999}. IPW-PLS is a wrapper method available within the \verb|plsVarSel| package \cite{Liland2023} and has relatively modest computational time requirements.
	\begin{gather}\label{eq:regpls}
		\begin{split}
			\textbf{Mod\stepcounter{mycounter}\themycounter}:\quad \widehat{\REC_t} &= 
			-\underset{(0.12866)}{0.00688632}\,{\CO}
			+\underset{(0.12157)^{***}}{0.424348}\,{\DINV}
			-\underset{(0.20940)}{0.0720370}\,{\EG}
			-\underset{(0.19641)}{0.254519}\,{\IND}\\
			&+\underset{(0.11321){*}}{0.208158}\,{\INFL}
			-\underset{(0.12269)^{*}}{0.241568}\,{\TR}
			+\underset{(0.11045)^{***}}{0.393072}\,{\URB}
		\end{split}
		\notag \\
		T = 32 \quad \bar{R}^2 = 0.6520 \quad F(7,24) = 9.2967^{***} \quad \hat{\sigma} = 0.58993 \quad  DW = 1.573470 \notag \\
		\centerline{(standard errors in parentheses; $^{*}p<0.1$, $^{**}p<0.05$, $^{***}p<0.01$).} \notag
	\end{gather}

	\subsection{Boruta}

	The Boruta algorithm is an all-relevant feature selection wrapper algorithm built around the random forest classification algorithm \cite{KursaetRudnicki2010}. Boruta takes features from the original dataset and creates a permuted copy of them by randomly shuffling each feature (the so-called ``shadow features''). Then, it runs a random forest classifier on these old and new feature sets to determine their importance using a statistical test. If the real feature has a higher importance than the maximum importance of shadow features, then the feature is considered significant and thus retained; otherwise, it is considered insignificant and is removed from the dataset in iteration. The algorithm is available within the \verb|Boruta| package \cite{KursaetRudnicki2023}.	
	\begin{gather}\label{eq:regboruta}
		\begin{split}
			\textbf{Mod\stepcounter{mycounter}\themycounter}:\quad \widehat{\REC_t} &= 
			\underset{(0.12010)^{**}}{0.323922}\,{\DINV_t}
			+\underset{(0.21296)}{0.119914}\,{\EG_t}
			+\underset{(0.17138)^{*}}{0.344490}\,{\FDI_t}
			-\underset{(0.18379)^{*}}{0.341631}\,{\IND_t}\\
			&+\underset{(0.10962)^{**}}{0.273522}\,{\INFL_t}
			-\underset{(0.16754)^{***}}{0.489998}\,{\TR_t}
			+\underset{(0.11634)^{**}}{0.281320}\,{\URB_t}
		\end{split}
		\notag \\
		T = 32 \quad \bar{R}^2 = 0.7021 \quad F(7,24) = 11.437^{***} \quad \hat{\sigma} = 0.54580 \quad DW = 1.356606\notag \\
		\centerline{(standard errors in parentheses; $^{*}p<0.1$, $^{**}p<0.05$, $^{***}p<0.01$).} \notag
	\end{gather}

	\subsection{Simulated annealing}

	This is a metaheuristic approach inspired by the annealing metallurgical process \cite{LaarhovenandAarts1987}. The simulated annealing algorithm takes an initial random subset of features and specifies the number of iterations. Then, for each iteration, a small perturbation is applied to the current feature subset (some features are randomly kept or removed), a model is fitted, and the predictive performance is calculated. If the performance improves, then the new subset is accepted and is updated as the current state. Otherwise, an acceptance probability is calculated, and a random number uniformly in the range $[0, 1]$ is generated. If the random number is greater than the acceptance probability, then the new feature set is rejected and the previous feature set is used. If the random number is larger than the acceptance probability, then the new subset is rejected and the current subset is kept; otherwise, the new subset is accepted and is updated as the current state.

	In our analysis, we used simulated annealing feature selection (\texttt{safs}) which is available within the \texttt{caret} package \cite{Kuhn2008, Kuhnetal2023}. 
	\begin{gather}\label{eq:regsa}
		\begin{split}
			\textbf{Mod\stepcounter{mycounter}\themycounter}:\quad \widehat{\REC_t} &= 
			-\underset{(0.13895)}{0.232931}\,{\CO_t}
			+\underset{(0.15359)^{***}}{0.702008}\,{\FDI_t}
			-\underset{(0.12545)}{0.167100}\,{\INC_t}\\
			& +\underset{(0.12478)^{*}}{0.327332}\,{\INFL_t}
			+\underset{(0.14071)}{0.130484}\,{\TOUR_t}
			-\underset{(0.15536)^{***}}{0.789605}\,{\TR_t}
		\end{split}
		\notag \\
		T = 32 \quad \bar{R}^2 = 0.5642 \quad F(6,25) = 7.6887^{***} \quad \hat{\sigma} = 0.66016 \quad DW = 0.717327 \notag \\
		\centerline{(standard errors in parentheses; $^{*}p<0.1$, $^{**}p<0.05$, $^{***}p<0.01$).} \notag
	\end{gather}
	
	\subsection{Genetic algorithms}
	Genetic algorithms (GAs) are heuristic search algorithms inspired by biological evolution, natural selection principles, and the survival of the fittest \cite{Leardietal1992,Mitchell1998}. Similar to evolution theory, a GA begins by initializing a population of potential solutions that represent various subsets of features in the context of feature selection. The quality of these feature subsets are assessed using a fitness function similar to natural selection in biology. Subsequently, GAs employ selection, crossover, and mutation to evolve the feature subsets over successive generations, with promising subsets having a higher chance of passing on to the next generation. Termination criteria determine when the GA’s search process concludes, and the best-performing feature subset is ultimately selected for use in constructing a machine learning model.

	We used genetic algorithm feature selection (\texttt{gafs}), also available within the \texttt{caret} package, which conducts a supervised binary search of the regressor space using a genetic algorithm \cite{Kuhn2008, Kuhnetal2023}.
	\begin{gather}\label{eq:regga}
		\begin{split}
			\textbf{Mod\stepcounter{mycounter}\themycounter}:\quad \widehat{\REC_t} &= 
			-\underset{(0.23704)^{**}}{0.604142}\,{\EG_t}
			+\underset{(0.16942)}{0.0771758}\,{\FDI_t}
			+\underset{(0.23947)}{0.320353}\,{\TOUR_t}
			+\underset{(0.16615)^{**}}{0.399564}\,{\URB_t}
		\end{split}
		\notag \\
		T = 32 \quad \bar{R}^2 = 0.3182 \quad F(4,27) = 4.6177^{***} \quad \hat{\sigma} = 0.82568 \quad  DW = 0.724634\notag \\
		\centerline{(standard errors in parentheses; $^{*}p<0.1$, $^{**}p<0.05$, $^{***}p<0.01$).} \notag
	\end{gather}

	\section{Summary}
	\subsection{Relative importance for linear regression}
\begin{figure}[htbp]
		\centering
		\caption{Bar plots of the relative importance for $\REC$ using lmg metrics ($R^2 = 82\%$, metrics are normalized to sum 100\%).}
		\includegraphics[width=0.90\linewidth]{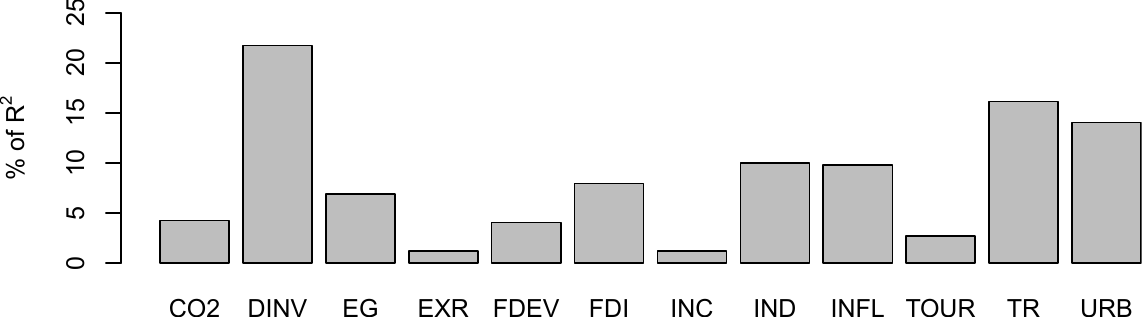}
		\label{fig:relaimpo}
	\end{figure}

	Here, we assess the relative importance of each regressor for the linear model in \eqref{eq:linreg}.  This method estimates the proportion of the linear model's R-squared ($ R^2 $) contributed by each individual regressor. We use the \texttt{relaimpo} package and use the ``lmg'' metrics which is a $ R^2 $ partitioned by averaging sequential sums of squares over all orderings of the regressors \cite{Groemping2006,GroempingMatthias2023}. As shown in Figure \ref{fig:relaimpo},  domestic investment ($\DINV$)  contributes the most to the linear model's $ R^2 $, followed by trade openness ($\TR$) and urbanization ($\URB$). Exchange rate ($\EXR$) has the least contribution among the regressors.

	We note that in our analysis, the features selected by the Boruta algorithm (\hyperref[eq:reglasso]{\textbf{Mod4}}) match the six most prominent features, as determined by the lmg metrics. These features, listed in order of importance, include domestic investment ($\DINV$), trade openness ($\TR$), urbanization ($\URB$), industry development ($\IND$), inflation ($\INFL$), and foreign direct investment ($\FDI$).

	\subsection{Evaluation of linear models}
	To evaluate the performance of each linear least squares model, we apply metrics such as adjusted R-squared ($\bar{R}^2$), Mallows' $C_p$, AIC, and BIC (see Table \ref{tab:evalmod}). For the LASSO model  (\hyperref[eq:reglasso]{\textbf{Mod4}}) particularly, we apply post-LASSO OLS, which involves performing ordinary least squares on the selected features \cite{BelloniChernozhukov2013}. In both terms of Mallows' $C_p$ and BIC, the best subset regression model  (\hyperref[eq:regbsr]{\textbf{Mod5}}) emerges as the top-performing model, with the key features being domestic investment ($\DINV$), financial development ($\FDEV$), foreign direct investment ($\FDI$), industrial development ($\IND$), inflation ($\INFL$), tourism development ($\TOUR$), and trade openness ($\TR$). These same features, along with environmental quality ($\CO$), are selected in the stepwise regression model (\hyperref[eq:regstep]{\textbf{Mod6}}), which is characterized by the lowest $\bar{R}^2$ and AIC values among all models. We note that in our results, the GA model (\hyperref[eq:regga]{\textbf{Mod11}}) is the least efficient model.

	\begin{table}[htbp]
		\centering
		\caption{Comparison of $\bar{R}^2$, Mallows' $C_p$, AIC, and BIC for all models.}
		\begin{tabular}{@{}lD{.}{.}{2.3}D{.}{.}{2.3}D{.}{.}{2.3}D{.}{.}{2.3}D{.}{.}{2.3}}
			\toprule
			Models & \multicolumn{1}{c}{Nb. of regressors} & \multicolumn{1}{c}{$\bar{R}^2$} & \multicolumn{1}{c}{Mallows' $C_p$} & \multicolumn{1}{c}{AIC } & \multicolumn{1}{c}{BIC} \\
			\midrule
			\hyperref[eq:regfull]{\textbf{Mod1}} & 12  & 0.7063 & 13.0000 & 62.9236 & 83.4439 \\
			\hyperref[eq:regcor]{\textbf{Mod2}} & 11  & 0.7205 & 11.0306 & 60.9750 & 80.0296 \\
			\hyperref[eq:reglasso]{\textbf{Mod4}} & 7   & 0.7315 & 5.9383 & 57.5250 & 70.7166 \\
			\hyperref[eq:regbsr]{\textbf{Mod5}} & 7   & 0.7360 & \mybold{5.5700} & 56.9833 & \mybold{70.1749} \\
			\hyperref[eq:regstep]{\textbf{Mod6}} & 8   & \mybold{0.7423} & 6.1809 & \mybold{56.8532} & 71.5105 \\
			\hyperref[eq:regrfe]{\textbf{Mod7}} & 8   & 0.7267 & 7.3997 & 58.7297 & 73.3870 \\
			\hyperref[eq:regpls]{\textbf{Mod8}} & 7   & 0.6520 & 12.4390 & 65.8299 & 79.0215 \\
			\hyperref[eq:regboruta]{\textbf{Mod9}} & 7   & 0.7021 & 8.3438 & 60.8544 & 74.0460 \\
			\hyperref[eq:regsa]{\textbf{Mod10}} & 6   & 0.5642 & 19.0973 & 72.3350 & 84.0609 \\
			\hyperref[eq:regga]{\textbf{Mod11}} & 4   & 0.3182 & 40.6754 & 85.1166 & 93.9110 \\
			\bottomrule
		\end{tabular}%
		\label{tab:evalmod}%
	\end{table}%

	\subsection{Models summary}
	Table \ref{tab:summary} provides a summary of the signs and statistical significance of the parameters associated with each model. Domestic investment ($\DINV$), foreign direct investment ($\FDI$), industrial development ($\IND$), price inflation ($\INFL$), and trade openness ($\TR$) are the most selected features that significantly impact renewable energy consumption ($\REC$) in Madagascar. Note that, with the exception of the PCR, LASSO, and IPW-PLS models (\hyperref[eq:regpcr]{\textbf{Mod2}}, \hyperref[eq:regpcr]{\textbf{Mod4}}, and \hyperref[eq:regpls]{\textbf{Mod8}}, respectively), $\TR$ consistently exhibits the largest coefficients among the other models.

	Financial development ($\FDEV$), tourism development ($\TOUR$), and urbanization ($\URB$) positively influence $\REC$. Conversely, environmental quality ($\CO$) and exchange rate ($\EXR$) have a negative impact on $\REC$. Economic growth ($\EG$) and income distribution ($\INC$) exhibit mixed effects depending on the model and do not display statistical significance.

	\setcounter{mycounter}{0}
	\begin{table}[htbp]
		\centering
		\caption{Models summary.}
		\begin{adjustbox}{width=\linewidth,center}
			\begin{tabular}{@{}lllllllllllll@{}}
				\toprule
				Models & $ \CO $ & $ \DINV $ & $ \EG $ & $ \EXR $ & $ \FDEV $ & $ \FDI $ & $ \INC $ & $ \IND $ & $ \INFL $ & $ \TOUR $ & $ \TR $ & $ \URB $ \\
				\midrule
				\hyperref[eq:regfull]{\textbf{Mod\stepcounter{mycounter}\themycounter}}  & $-$   & $+^{**}$ & $+$   & $-$   & $+^{**}$ & $+^{**}$ & $-$   & $-$   & $+^{*}$ & $+$   & $-^{***}$ & $+$ \\
				\hyperref[eq:regpcr]{\textbf{Mod\stepcounter{mycounter}\themycounter}} & $-$   & $+$   & $-$   & $-$   & $+$   & $+$   & $+$   & $-$   & $+$   & $+$   & $-$   & $+$ \\
				\hyperref[eq:regcor]{\textbf{Mod\stepcounter{mycounter}\themycounter}}  & $-$   & $+^{***}$ &       & $-$   & $+^{**}$ & $+^{**}$ & $-$   & $-^{*}$ & $+^{**}$ & $+$   & $-^{***}$ & $+$ \\
				\hyperref[eq:reglasso]{\textbf{Mod\stepcounter{mycounter}\themycounter}} &       & $+$   &       &       & $+$   & $+$   &       & $-$   & $+$   &       & $-$   & $+$ \\
				\hyperref[eq:regbsr]{\textbf{Mod\stepcounter{mycounter}\themycounter}}  &       & $+^{***}$ &       &       & $+^{***}$ & $+^{***}$ &       & $-^{**}$ & $+^{**}$ &  $+$     & $-^{***}$ &  \\
				\hyperref[eq:regstep]{\textbf{Mod\stepcounter{mycounter}\themycounter}} & $-$   & $+^{***}$ &       &       & $+^{***}$ & $+^{***}$ &       & $-^{*}$ & $+^{**}$ & $+^{*}$ & $-^{***}$ &  \\
				\hyperref[eq:regrfe]{\textbf{Mod\stepcounter{mycounter}\themycounter}}  &       & $+^{***}$ &   $+$    &       & $+^{*}$ & $+^{**}$ &       & $-^{*}$ & $+^{**}$ &    & $-^{***}$ &  \\
				\hyperref[eq:regpls]{\textbf{Mod\stepcounter{mycounter}\themycounter}} & $-$   & $+^{***}$ & $-$   &       &       &       &       & $-$   & $+^{*}$ &       & $-^{*}$ & $+^{***}$ \\
				\hyperref[eq:regboruta]{\textbf{Mod\stepcounter{mycounter}\themycounter}} &       & $+^{**}$ & $+$   &       &       & $+^{*}$ &       & $-^{*}$ & $+^{**}$ &       & $-^{***}$ & $+^{**}$ \\
				\hyperref[eq:regsa]{\textbf{Mod\stepcounter{mycounter}\themycounter}} & $-$   &       &       &       &       & $+^{***}$ & $-$   &       & $+^{**}$ & $+$   & $-^{***}$ &  \\
				\hyperref[eq:regga]{\textbf{Mod\stepcounter{mycounter}\themycounter}} &    &       &  $-^{**}$     &       &       & $+^{}$ &   &       &  & $+$   &  &  $+^{**}$\\
				\bottomrule
				\multicolumn{10}{l}{$^{*}p<0.1$, $^{**}p<0.05$, $^{***}p<0.01$}
			\end{tabular}%
		\end{adjustbox}
		\label{tab:summary}%
	\end{table}%

	On the basis of the results in Table \ref{tab:summary}, we infer the following remarks.
	\begin{enumerate}[label=(\roman*)]
		
		\item Increases in greenhouse gas emissions are mainly due to unsustainable energy practices, such as dependance on non-renewable energy sources and overdependence on fossil fuels for transport. Therefore, the corresponding negative impact can be clearly observed.
		
		\item The positive impact of domestic and foreign investment shows the effectiveness of investment in renewable energy generation. It also reflects the government and private sector’s commitment to promote clean energy sources.
		
		\item The mixed signs associated with $\EG$ indicate that a full energy transition has not yet been achieved during economic growth. It is important to recognize that countries often use a combination of energy sources during periods of economic growth, and the extent to which they transition to renewable energy can vary depending on several factors, including policy, technology, and availability of resources.
		
		\item Exchange rates negatively impact the adoption of renewable energy sources, particularly by affecting the prices of imported equipment. If the exchange rate increases the cost of imported renewable energy equipment, it may discourage their adoption because higher costs may limit the affordability of such technologies. Exchange rate fluctuations also create uncertainty in energy planning and can lead to a lack of consistency in renewable energy policies and projects.
		
		\item The two-fold impact of income distribution reflects consumer choice of energy sources. When individuals experience an increase in income, they have two options: they may choose polluting energy sources that are often more affordable and accessible, thereby reducing their immediate costs, or they may see an opportunity to invest in more sustainable renewable energy sources.
		
		\item Considering the negative impact of industry development, it is clear that industry-initiated energy transition is still ongoing and is not yet sufficiently observable on the basis of our data. This result also suggests that industrial development significantly increases energy demand, in particular in energy-intensive sectors such as manufacturing, which may lead to a greater dependence on conventional energy sources such as fossil fuels.
		
		\item Rising global oil prices have led to increased energy prices, which is one of the main inflation drivers in Madagascar. Therefore, individuals and businesses are more likely to invest in renewable energy technologies to reduce costs and take advantage of government incentives.
		
		\item The positive effect of financial development indicates that the availability of funding programs and credits to promote renewable energy sources is effective.
		
		\item The influx of tourists to Madagascar has driven investment in sustainable energy solutions. Therefore, tourism is an efficient catalyst for the adoption of sustainable energy in Madagascar because it raises awareness of renewable energy and creates demand for environmentally responsible practices.
		
		\item The magnitude and negative sign of the coefficient associated with the trade openness variable indicates the extent to which Madagascar’s economy relies on imports to meet its energy requirements. In terms of exports, this can be explained by the fact that the infrastructure and transport systems used for the production of export goods, such as minerals, are not yet fully aligned with energy-efficient or renewable energy technologies.
		
		\item Urbanisation in Madagascar has a positive impact on the adoption of renewable energy sources, mainly due to the increasing demand for electricity in urban areas. Because of the challenges associated with frequent power cuts, the urban population, the government, and the private sector have begun to seek sustainable solutions, with increasing emphasis on renewable energy sources.
		
	\end{enumerate}


\begin{thebibliography}{99}
		
		\bibitem{AfDB2023}
		African Development Bank (AfDB), ``Madagascar Economic Outlook'', Retrieved October 10, 2023 from \url{https://www.afdb.org/en/knowledge/publications/african-economic-outlook}
		
		
		\bibitem{ARCEB2022} 
		Projet d'Appui au Renforcement des Capacités d'Analyse des Facteurs de Vulnérabilité Structurelle et la Promotion de l'Économie Bleue (ARCEB), \href{http://www.mef.gov.mg/assets/vendor/ckeditor/plugins/kcfinder/upload/files/ARCEB/Etude\%20Secteur\%20Energie\%20Rapport\%20Final\%20_\%20Version\%20finale\%20Novembre\%202022.pdf}{\textit{Analyse Approfondie de l’Ensemble des Contraintes qui Freinent le Développement du Secteur de l’Électricité}}, Ministère de l’Énergie et des Hydrocarbures - Projet ARCEB, Madagascar, 2022.
		
		
		\bibitem{AdusahPoku2016}
		F.\ Adusah-Poku, ``\href{https://doi.org/10.18488/journal.82/2016.3.1/82.1.1.16}{Carbon dioxide emissions, urbanization and population: Empirical evidence in Sub-Saharan Africa}'', \textit{Energy Economics Letters} \textbf{3} (2016), 1--16. 
		
		
		\bibitem{Ansarietal2021} 
		M.\ A.\ Ansari, S.\ Haider \& T.\ Masood, ``\href{https://doi.org/10.1007/s11356-020-10786-0}{Do renewable energy and globalization enhance ecological footprint: an analysis of top renewable energy countries?}'', \textit{Environmental Science and Pollution Research} \textbf{28} (2021), 6719--6732.
		
		
		\bibitem{Azam2016}
		M.\ Azam, A.\ Qayyum Khan, E.\ Zafeiriou \& G.\ Arabatzis,
		``\href{https://doi.org/10.1016/j.rser.2015.12.082}{Socio-economic determinants of energy consumption: An empirical survey for Greece}'', \textit{Renewable and Sustainable Energy Reviews} \textbf{57} (2016), 1556--1567.
		
		
		\bibitem{BamatiandRaoofi2020} 
		N.\ Bamati \& A.\ Raoofi,	``\href{https://doi.org/10.1016/j.renene.2019.11.098}{Development level and the impact of technological factor on renewable energy production}'', \textit{Renewable Energy} \textbf{151} (2020), 946-955.
		
		
		\bibitem{BFM2023a} 
		Banky Foiben'i Madagasikara, ``Bulletin de la BFM''. Retrieved July 01, 2023, from \url{https://www.banky-foibe.mg/vente_bulletin-de-bfm}
		
		
		\bibitem{BFM2023b} 
		Banky Foiben'i Madagasikara, ``Marché de Changes''. Retrieved July 01, 2023, from \url{https://www.banky-foibe.mg/marche_marche-de-change\#recherche-taux-de-change}
		
		
		\bibitem{BFM2023c} 
		Banky Foiben'i Madagasikara, ``Rapport Annuel''. Retrieved July 01, 2023, from \url{https://www.banky-foibe.mg/_rapport-annuel}
		
		
		\bibitem{Banoetal2021}
		S.\ Bano, M.\ Alam, A. Khan \& L.\ Liu, ``\href{https://doi.org/10.1007/s10668-021-01275-6}{The nexus of tourism, renewable energy, income, and environmental quality: an empirical analysis of Pakistan}''. \textit{Environ.\ Dev.\ Sustain}.\ \textbf{23} (2021), 14854--14877. 
		
		
		\bibitem{BelloniChernozhukov2013}
		A.\ Belloni \& V.\ Chernozhukov, ``\href{https://doi.org/10.3150/11-BEJ410 }{Least squares after model selection in high-dimensional sparse models}'', \textit{Bernoulli} \textbf{19} (2013), 521--547. 
		
		
		\bibitem{BenJebli2019}
		M.\ Ben Jebli, S.\ Ben Youssef \& N.\ Apergis, ``\href{https://doi.org/10.1186/s40503-019-0063-7}{The dynamic linkage between renewable energy, tourism, CO2 emissions, economic growth, foreign direct investment, and trade}'', \textit{Lat.\ Am.\ Econ.\ Rev}.\ \textbf{28} (2019), 1--19.
		
		
		\bibitem{Casseetal2022}
		T.\ Casse, M.\ H.\ Razafintsalama \& A.\ Milhøj, ``\href{https://doi.org/10.1007\%2Fs43545-021-00309-0}{A trade-off between conservation, development, and tourism in the vicinity of the Andasibe-Mantadia National Park, Madagascar}'', \textit{SN Soc Sci} \textbf{2} (2022), 1--28.
		
		
		\bibitem{Cieltextile2022}
		Ciel Textile, ``Sustainability report 2022. Winning Well''. Retrieved October 10, 2023 from \url{https://cieltextile.com/Sustainability-Report-2022.pdf}		
		
		
		\bibitem{Benli2021}
		M.\ Benli, ``\href{https://dergipark.org.tr/en/pub/bsbd/issue/67912/996509}{Foreign direct investment and energy consumption in developing economies: An Analysis of heterogeneous dynamic panel data models with cross sectional dependency}'', \textit{Balkan Sosyal Bilimler Dergisi} \textbf{10} (2021), 18--27.
		
		
		\bibitem{ChenandJeong2007}
		X.\ Chen \& J.\ C.\ Jeong, ``\href{https://doi.org/10.1109/ICMLA.2007.35}{Enhanced recursive feature elimination}'', \textit{Sixth International Conference on Machine Learning and Applications -- ICMLA} (2007),  429--435.
		
		
		\bibitem{Clark2012}
		M.\ Clark, ``\href{https://www.american.edu/cas/economics/ejournal/upload/Clark_accessible.pdf}{Deforestation in Madagascar: Consequences of population growth and unsustainable agricultural processes}'', \textit{Global Majority E-Journal} \textbf{3} (2012), 61--71.
		
		
		\bibitem{Darwinetal2022}
		R.\ Darwin, D.\ W.\ Sari \& U.\ Heriqbaldi, ``\href{https://doi.org/10.32479/ijeep.13552}{Dynamic linkages between energy consumption, foreign direct investment, and economic growth: A New insight from developing countries in Asia}'', \textit{International Journal of Energy Economics and Policy} \textbf{12} (2022), 30--36.
		
		
		\bibitem{DCRPPresidence2021}
		Direction de la Communication et des Relations Publiques, ``Signature des contrats du projet Sahofika et PRIRTEM 2'', \textit{Présidence de la République de Madagascar} (November 15, 2021). Retrieved October 01, 2023, from \url{https://www.presidence.gov.mg/actualites/1443-signature-des-contrats-du-projet-sahofika-et-prirtem-2.html}
		
		
		\bibitem{DCRPPresidence2023}
		Direction de la Communication et des Relations Publiques, ``Palais d'Etat d'Iavoloha: Signature du Projet Volobe Amont'', \textit{Présidence de la République de Madagascar} (May 26, 2023). Retrieved October 01, 2023, from \url{https://www.presidence.gov.mg/actualites/1896-palais-d-etat-d-iavoloha-signature-du-projet-volobe-amont.html}
		
		\bibitem{DSCM2019} 
		Direction des Statistiques des Conditions de Vie des Ménages, \textit{Le nouvel indice des prix à la consommation}, Institut National de la Statistique, Madagascar, 2019.
		
		
		\bibitem{Douane2023}
		Douane Malagasy, ``Commerce exterieur et exoneration de droits et taxes à l'importation 2023''. Retrieved August 01, 2023, from \url{http://www.douanes.gov.mg/stat/commerce-exterieur-et-exoneration-de-droits-et-taxes-a-limportation-2023/}
		
		
		\bibitem{DutoitMcLachlan}
		Du Toit McLachlan, ``Madagascar'', \textit{ISS African Futures} (June 8, 2023), Retrieved  September 10, 2023, from \url{https://futures.issafrica.org/geographic/countries/madagascar/}
		
		
		\bibitem{EDBM2021}
		Economic Development Board of Madagascar (EDBM), ``\href{https://edbm.mg/wp-content/uploads/2021/08/Yearbook-economique-Madagascar-2021.pdf}{Une zone industrielle dédiée au textile}'', in \textit{Yearbook Rapport Economique. Madagascar 2021, L'Emergence Malagsy}, EDBM, 2021.
		
		
		\bibitem{Effiong2016}
		E.\ Effiong, ``\href{https://mpra.ub.uni-muenchen.de/73224/}{Urbanization and environmental quality in Africa}'', \textit{Munich Personal RePEc Archive} \textbf{73224}, University Library of Munich, Germany, 2016.
		
		
		\bibitem{Forina1999}
		M. Forina, C. Casolino \& C. Pizarro Millan, ``\href{https://doi.org/10.1002/(SICI)1099-128X(199903/04)13:2\%3C165::AID-CEM535\%3E3.0.CO;2-Y}{Iterative predictor weighting (IPW) PLS: a technique for the elimination of useless predictors in regression problems}'', \textit{Journal of Chemometrics} \textbf{13} (1999), 165--184.
		
		
		\bibitem{Foye2023}
		V.\ O.\ Foye, ``\href{https://doi.org/10.1016/j.totert.2022.100022}{Macroeconomic determinants of renewable energy penetration: Evidence from Nigeria}'', \textit{Total Environment Research Themes} \textbf{5} (2023), 1--13.
		
		
		\bibitem{GeladiKowalski1986}
		P.\ Geladi \& B.\ R.\ Kowalski, ``\href{https://doi.org/10.1016/0003-2670(86)80028-9}{Partial least-squares regression: a tutorial}'',	\textit{Analytica Chimica Acta} \textbf{185} (1986), 1--17.
		
		
		\bibitem{Gossling2010}
		Stefan Gössling, ``\href{https://doi.org/10.1080/09669580008667376}{Sustainable tourism development in developing countries: Some aspects of energy use}'', \textit{Journal of Sustainable Tourism} \textbf{8} (2010), 410--425.
		
		
		\bibitem{Gozgoretal2020}
		G.\ Gozgor, M.\ Kumar Mahalik, E.\ Demir \& H.\ Padhan,
		``\href{https://doi.org/10.1016/j.enpol.2020.111365}{The impact of economic globalization on renewable energy in the OECD countries}'', \textit{Energy Policy} \textbf{139} (2020), 1--13.
		
		
		\bibitem{Groemping2006} 
		U.\ Groemping, ``\href{https://doi.org/10.18637/jss.v017.i01}{Relative importance for linear regression in R: The package relaimpo}'', \textit{Journal of Statistical Software} \textbf{17} (2006), 1--27.
		
		
		\bibitem{GroempingMatthias2023} 
		U.\ Groemping \& L. Matthias, \textit{relaimpo: Relative Importance of Regressors in Linear Models},  R package 2.2-6, 2023. Available for download at {\tt\url{https://cran.r-project.org/package=relaimpo}}
		
		
		\bibitem{GrossetandNguyen-Van} 
		F.\ Grosset \& P.\ Nguyen-Van, ``\href{ https://doi.org/10.3917/med.176.0025}{Consommation d’énergie et croissance économique en Afrique subsaharienne}'', \textit{Mondes en développement} \textbf{176}, 25--42.
		
		
		\bibitem{Guyonetal2002}
		I.\ Guyon, J.\ Weston, S.\ Barnhill, \& V. Vapnik, ``\href{https://doi.org/10.1023/A:1012487302797}{Gene selection for cancer classification using support vector machines}'', \textit{Machine Learning} \textbf{46} (2022), 389--422. 
		
		
		\bibitem{HafalianavalonaExpress2017}
		S.\ M.\ Hafalianavalona, ``Développement -- L'énergie verte en soutien au tourisme'', \textit{L'express de Madagascar} (June 22, 2017). Retrieved Saptember 10, 2023, from \url{https://lexpress.mg/22/06/2017/developpement-lenergie-verte-en-soutien-au-tourisme/}
		
		
		\bibitem{INSTATCN2022b}
		Institut National de la Statistique (INSTAT), ``Comptes Nationaux'', \textit{Wayback Machine} (December 28, 2021) Retrieved August 10, 2023, from \url{https://web.archive.org/web/20211228223849/https://www.instat.mg/statistiques/bases-de-donnees/comptes-nationaux}
		
		
		\bibitem{INSTATIPC2022a}
		Institut National de la Statistique (INSTAT), ``Indice des prix à la Consommation (IPC)'', \textit{Wayback Machine} (January 20, 2022). Retrieved August 10, 2023, from  \url{https://web.archive.org/web/20220120074937/https://www.instat.mg/statistiques/bases-de-donnees/nipc}	
		
		
		\bibitem{INSTATCN2022c}
		Institut National de la Statistique (INSTAT), ``Tableau de Bord de l'Économie'', \textit{Wayback Machine} (September 28, 2022)  Retrieved August 10, 2023, from \url{https://web.archive.org/web/20220928222805/http://www.instat.mg/thematique/economie}	
		
		
		\bibitem{Iqbaletal2023}
		S.\ Iqbal, Y.\ Wang, S.\ Ali, M.\ A.\ Haider \& N.\ Amin, ``\href{https://doi.org/10.1016/j.renene.2022.11.071}{Shifting to a green economy: Asymmetric macroeconomic determinants of renewable energy production in Pakistan}'', \textit{Renewable Energy} \textbf{202} (2023), 234--241.
		
		
		\bibitem{Islametal2022}
		Md.\ M.\ Islam, M.\ Irfan, M.\ Shahbaz \& X.\ V.\ Vo, ``\href{https://doi.org/10.1016/j.renene.2021.12.020}{Renewable and non-renewable energy consumption in Bangladesh: The relative influencing profiles of economic factors, urbanization, physical infrastructure and institutional quality}'', \textit{Renewable Energy} \textbf{184} (2022), 1130--1149.
		
		
		\bibitem{JIRAMA2023}
		Jiro sy Rano Malagasy (JIRAMA), ``Consommation d'électricité par branches d'activités (2007--2022)'' [unpublished raw data], \textit{JIRAMA}, Madagascar.
		
		
		\bibitem{Knowledge4Policy}
		Knowledge for Policy, ``République de Madagascar - Programme indicatif pluriannuel 2021-2027''. Retrieved October 10, 2023, from  \url{https://knowledge4policy.ec.europa.eu/file/republique-de-madagascar-programme-indicatif-pluriannuel-2021-2027_en}
		
		
		\bibitem{Kuhn2008}
		M.\ Kuhn, ``\href{https://doi.org/10.18637/jss.v028.i05}{Building predictive models in R using the caret package}'', \textit{Journal of Statistical Software} \textbf{28} (2008), 1--26.
		
		
		\bibitem{Kuhnetal2023}
		M.\ Kuhn, J.\ Wing, S.\ Weston, A.\ Williams, C.\ Keefer, A.\ Engelhardt, T.\ Cooper, Z.\ Mayer, B.\ Kenkel, R Core Team, M.\ Benesty, R.\ Lescarbeau, A.\ Ziem, L.\ Scrucca, Y.\ Tang, C.\ Candan \& T.\ Hunt, c\textit{aret: Classification and Regression Training}, R package 6.0-94, 2023. Available for download at \url{https://cran.r-project.org/package=caret}
		
		
		\bibitem{KursaetRudnicki2010}
		M.\ B.\ Kursa \&  W.\ R.\ Rudnicki,  ``\href{https://doi.org/10.18637/jss.v036.i11}{Feature selection with the Boruta package}'', \textit{Journal of Statistical Software} \textbf{36} (2010), 1--13. 
		
		
		\bibitem{KursaetRudnicki2023}
		M.\ B.\ Kursa \&  W.\ R.\ Rudnicki, \textit{Boruta: Wrapper Algorithm for All Relevant Feature Selection}, R package 8.0.0., 2023. Available for download at \url{https://cran.r-project.org/package=Boruta}
		
		
		\bibitem{LaarhovenandAarts1987}
		P.\ J.\ M. van Laarhoven  \& E.\ H.\ L.\ Aarts, ``\href{https://doi.org/10.1007/978-94-015-7744-1_2}{Simulated annealing}'', in \textit{Simulated Annealing: Theory and Applications, Simulated annealing, Mathematics and Its Applications} \textbf{37} (1987), Springer, pp.\ 7--15.	
		
		
		\bibitem{Leardietal1992} 
		R.\ Leardi, R.\ Boggia \& M.\ Terrile , ``\href{https://doi.org/10.1002/cem.1180060506}{Genetic algorithms as a strategy for feature selection}'', \textit{Journal of Chemometrics} \textbf{6} (1992), 267--281. 
		
		
		\bibitem{Lanetal2019} 
		J.\ Lan, M.\ Salmeri, B.\ Curnier \& Y.\ L.~Hamden, \href{https://greenminigrid.afdb.org/sites/default/files/gmg_madagascar-2.pdf}{\textit{Mini-Grid Market Opportunity Assessment: Madagascar}}, Green Mini-Grids Market Development Programme Document Series, African Development Bank Group, 2019.
		
		\bibitem{Liland2023}
		K.\ H.\ Liland, T. Mehmood \& S.\ S\ae{}b\o{}, \textit{plsVarSel: Variable Selection in Partial Least Squares}, R package 8.0.0, 2023. Available for download at \url{https://cran.r-project.org/package=plsVarSel}
		
		
		\bibitem{Maalejetal2018}
		R.\ Maalej with contributions from B.\ Toorabally, T.\ Woodward, L.\ Rasolofomanana, R.\ Randrianarisoa, R. Rakoto-Harisoa, H.\ Northover, J.\ Garrett \& S. Kempster, \href{https://washmatters.wateraid.org/publications/mineral-rights-to-human-rights}{\textit{Mineral rights to human rights: mobilising resources from the extractive industries for water, sanitation and hygien. Case study: Madagascar}}, WaterAid and Moore Stephens LLP, 2018.
		
		
		\bibitem{MalagasyNews2021}
		Malagasy News, ``Crédit Sunref: Une enveloppe de six millions d’euros pour les projets écoresponsables'', \textit{Malagasy News} (April 13, 2021). Retrieved October 10, 2023, from \url{https://malagasynews.com/economie/10340/}
		
		
		\bibitem{Maliketal2014}
		I.\ A.\ Malik, Ghamz-e-Ali Siyal, A.\ B.\ Abdullah, A.\ Alam, K.\ Zaman, P.\ Kyophilavong, M.\ Shahbaz, S.\ U.\ Baloch \& T.\ Shams, ``\href{https://doi.org/10.1016/j.rser.2014.05.090}{Turn on the lights: Macroeconomic factors affecting renewable energy in Pakistan}'', \textit{Renewable and Sustainable Energy Reviews} \textbf{38} (2014), 277--284.
		
		
		\bibitem{Mehmoodetal2012}
		T.\ Mehmood, K.\ H.\ Liland, L.\ Snipen \& S.\ S\ae{}b\o{}, ``\href{https://doi.org/10.1016/j.chemolab.2012.07.010}{A review of variable selection methods in Partial Least Squares Regression}'', \textit{Chemometrics and Intelligent Laboratory Systems} \textbf{118} (2012), 62--69.
		
		
		\bibitem{Mehmoodetal2022}
		T.\ Mehmood, S.\ S\ae{}b\o{} \& K.\ H.\ Liland, ``\href{https://doi.org/10.1002/cem.3226}{Comparison of variable selection methods in Partial Least Squares Regression}'',  \textit{Journal of Chemometrics} \textbf{34} (2020), 1-14.
		
		
		\bibitem{MEEF2018}
		Ministère de l'Environnement, de l'Écologie et des Forêts (MEEH), \href{https://www.forestcarbonpartnership.org/system/files/documents/strategie_nationale_redd_madagascar_final_13-06-18_accentue_0_1.pdf}{Stratégie Nationale REDD+ Madagascar}, MEEF - BNC-REDD+, Madagascar, 2018.
		
		
		\bibitem{MEH2015} 
		Ministère de l’Énergie et des Hydrocarbures (MEH), \href{http://www.ore.mg/Publication/Rapports/LettreDePolitique.pdf}{\textit{Lettre de Politique de l'Énergie de Madagascar 2015--2030}}, MEH, Madagascar, 2015.
		
		
		\bibitem{MEH2015a} 
		Ministère de l'Énergie et des Hydrocarbures (MEH), \href{http://www.ore.mg/Publication/Rapports/NouvellePolitiqueDel%27Energie.pdf}{\textit{Assistance pour le Développement d’une Nouvelle Politique de l’Énergie et d’une Stratégie pour la République de Madagascar – Phases 2 et 3. Document d’Étude de la Politique et Stratégie de l’Énergie}}, MEH, Madagascar, 2015.
		
		
		\bibitem{MEH2021} 
		Ministère de l’Énergie et des Hydrocarbures (MEH), \href{https://meh.mg/wp-content/uploads/2022/07/R5-Planification-Geospatiale-Rapport-final-V12.pdf}{\textit{Analyse des options d'électrification géospatiale au moindre coût pour un déploiement sur réseau et hors réseau}}, MEH, Madagascar, 2021
		
		
		\bibitem{MESUPRES2015}
		Ministère de l'Enseignement Supérieur et de la Recherche Scientifique (MESUPRES), \href{https://mg.chm-cbd.net/implementation/Documents_nationaux/document-cadre/plan-directeur-de-la-recherche/plan-directeur-de-la-recherche-sur-les-energies-renouvelables-2014-2018}{\textit{Plan Directeur de la Recherche sur les Énergies Renouvelables 2014-2018}}, MESUPRES - UNESCO, Madagascar, 2015.
		
		
		\bibitem{MinWEH2018}
		Ministry of Water, Energy and Hydrocarbons -- Scaling Renewable Energy Program (SREP), \href{https://www.cif.org/sites/cif_enc/files/srepinvestment_plan_for_madagascar_final.pdf}{\textit{Investment Plan for Renewable Energy in Madagascar}}, Ministry of Water, Energy and Hydrocarbons, Madagascar, 2018.
		
		
		\bibitem{Mitchell1998}
		M.\ Mitchell, \textit{An Introduction to Genetic Algorithms, MIT Press}, 1998.
		
		
		\bibitem{MohamedYusoffetal2023}
		N.\ Y.\ Mohamed Yusoff, A.\ R.\ Ridzuan, T.\ Soseco, Wahjoedi, B.\ S.\ Narmaditya \& L.C. Ann, ``\href{https://doi.org/10.3390/su15053891}{Comprehensive outlook on macroeconomic determinants for renewable energy in Malaysia}'', \textit{Sustainability} \textbf{15} (2023), 1--15. 
		
		
		\bibitem{MurshedAlam2021}
		M.\ Murshed \& M.\ S.\ Alam, ``\href{https://doi.org/10.1007/s11356-021-12516-6}{Estimating the macroeconomic determinants of total, renewable, and non-renewable energy demands in Bangladesh: the role of technological innovations}'', \textit{Environ. Sci. Pollut. Res}.\ \textbf{28} (2021), 30176--30196. 
		
		
		\bibitem{NaidooandLoot2020} 
		K.\ Naidoo \& C.\ Loot, \href{https://www.uncdf.org/article/6474/energy-and-the-poor-unpacking-the-investment-case-for-clean-energy}{\textit{Madagascar. Energy and the Poor: Unpacking the Investment Case for Clean Energy}}, UN Capital Development Fund (UNCDF), 2020.
		
		
		\bibitem{OMDF2020}
		Off-Grid Market Development Fund (OMDF), ``The opportunities of the off-grid solar market in Madagascar''. Retrieved September 30, 2023, from \url{https://omdf.mg/public/uploads/2020/11/Fr-OMDF-press-release-2020-11-11-final.pdf}
		
		
		\bibitem{OMH2021}
		Office Malgache des Hydrocarbures (OMH), ``Bulletin Pétrolier (Le Marché Pétrolier au 1er, 2ème, 3ème, 4ème trimestre 2021)''. Retrieved August 10, 2023, from  \url{http://www.omh.mg/index.php?idm=5&CL=bulpetro}
		
		
		\bibitem{OMH2023}
		Office Malgache des Hydrocarbures (OMH), ``Ventes nationales de produits pétroliers (2016--2022)'' [Unpublished raw data],  \textit{OMH}, Madagascar.
		
		
		\bibitem{OWID2023}
		Our World in Data, ``Madagascar: CO2 Country Profile''. Retrieved August 10, 2023, from \url{https://ourworldindata.org/co2/country/madagascar}
		
		
		\bibitem{Polat2018}
		B.\ Polat, ``\href{https://doi.org/10.19168/jyasar.340938}{The influence of FDI on energy consumption in developing and developed countries: A dynamic panel data approach}'', \textit{Journal of Yasar University} \textbf{13} (2018), 33--42.
		
		
		\bibitem{Praeneetal2017}
		J.\ P.\ Praene, M.\ H.\ Radanielina, V.\ R.\ Rakotoson, A.\ L.\ Andriamamonjy, F.\ Sinama, D.\ Morau \& H.\ T.\ Rakotondramiarana, ``\href{https://doi.org/10.1016/j.rser.2017.03.125}{Electricity generation from renewables in Madagascar: Opportunities and projections}'', \textit{Renewable and Sustainable Energy Reviews} \textbf{76} (2017), 1066--1079.
		
		
		\bibitem{Praeneetal2021}
		J.\ P.\ Praene, R.\ M.\ Rasamoelina \& L.\ Ayagapin,
		``\href{https://doi.org/10.1016/j.rser.2021.111321}{Past and prospective electricity scenarios in Madagascar: The role of government energy policies}'', \textit{Renewable and Sustainable Energy Reviews} \textbf{149} (2021), 1--12.
		
		
		\bibitem{PRC-ELEC2019}
		Programme de Révision du Cadre juridique du secteur Electricité (PRC-ELEC), appuyé par la Deutsche Gesellschaft für Internationale Zusammenarbeit (GIZ) GmbH, \href{http://www.ore.mg/Publication/Rapports/2019_Madagascar_avantages\%20fiscaux\%20EnR\%20(GIZ).pdf}{\textit{Guide sur les avantages fiscaux en faveur des énergies renouvelables}}, Ministère de l’Eau, de l’Énergie et des Hydrocarbures (MEEH), Office de Régulation de l’Électricité (ORE), Agence de Développement de l’Électrification Rurale (ADER), Madagascar, 2019.
		
		
		\bibitem{PERER2020}
		Promotion de l'Électrification par les Énergies Renouvelables (PERER), \href{https://edbm.mg/wp-content/uploads/2021/06/Manuel-financement-energie-2020.pdf}{\textit{Financement des projets d'énergies renouvelables à Madagascar}}, Deutsche Gesellschaft für Internationale Zusammenarbeit (GIZ) GmbH – Programme ``Promotion de l’Électrification par les Énergies Renouvelables'', 2020.
		
		
		\bibitem{Ramaharo2023}
		F.\ M.\ Ramaharo, ``Madagascar: International Tourism, Number of Arrivals'' \textit{Open Science Framework} (July 20, 2023). Retrieved August 10, 2023, from \url{https://doi.org/10.17605/OSF.IO/R6JHC}
		
		
		\bibitem{Ramaharo2023data}
		F.\ M.\ Ramaharo,
		Machine learning approach for assessing the determinants of renewable energy consumption in Madagascar, \textit{Open Science Framework} (October 27, 2023). Retrieved October 27, 2023, from \url{https://doi.org/10.17605/OSF.IO/XDAU4}
		
		
		\bibitem{Ramaharoetal2023}
		F.\ M.\ Ramaharo, E.\ M.\ Aljaona, F.\ K.\ Manitriniaina \& F.\ M.\ Randriamifidy, ``Madagascar: Cours indicatifs des principales devises'', \textit{Open Science Framework} (July 20, 2023). Retrieved August 10, 2023, from \url{https://doi.org/10.17605/OSF.IO/UFZ7V}
		
		
		\bibitem{RamaharoetRajaonarison2023}
		F.\ M.\ Ramaharo \& N.\ R.\ Rajaonarison, ``\href{https://mpra.ub.uni-muenchen.de/id/eprint/116142}{Principal component regression analysis of electricity consumption factors in Madagascar}'', \textit{Munich Personal RePEc Archive} \textbf{16142}, University Library of Munich, Germany, 2023.
		
		
		\bibitem{RamaharoRazanajatovo2023}
		F.\ M.\ Ramaharo \& Y.\ H.\ M.\ Razanajatovo, ``Rapport Economique et Financier'', \textit{Open Science Framework} (July 17, 2023). Retrieved August 20, 2023, from \url{https://doi.org/10.17605/OSF.IO/7RDBH} 
		
		
		\bibitem{Randrianariveloetal2022}
		M.\ N.\ R.\ Randrianarivelo, R.\ Rakotosaona, E.\ R. Andrianarison, \href{http://dx.doi.org/10.52155/ijpsat.v34.2.4621}{Changement climatique et transition enérgétique: Quel avenir pour Madagascar?}, \textit{International Journal of Progressive Sciences and Technologies} \textbf{34} (2022), 295--308.
		
		
		\bibitem{RasamoelinaPraene2018}
		R.\ M.\ Rasamoelina \& J.\ P.\ Praene, ``\href{https://www.researchgate.net/publication/327673404_Situation_energetique_de_Madagascar_Scenarios_bases_sur_la_decomposition_LMDI}{Situation énergétique de Madagascar: Scénarios basés sur la décomposition LMD}'', in \textit{Journées de Recherche des ISTs de Madagascar: Antananarivo}, 2018.
		
		
		\bibitem{Ritchieetal2020}
		H.\ Ritchie, M.\ Roser and P.\ Rosado, ``CO2 and Greenhouse Gas Emissions'', Our World in Data. Retrieved August 01, 2023, from \url{https://ourworldindata.org/co2-and-greenhouse-gas-emissions}
		
		
		\bibitem{Samouretal2022}
		A.\ Samour, M.\ M.\ Baskaya \& T.\ Tursoy, ``\href{https://doi.org/10.3390/su14031208}{The impact of financial development and FDI on renewable energy in the UAE: A path towards sustainable development}'', \textit{Sustainability} \textbf{14} (2022), 1--14. 
		
		
		\bibitem{Shahetal2022}
		M.\ H.\ Shah, I.\ Ullah, S. Salem, S. Ashfaq, A. Rehman, M.\ Zeeshan \& Z.\ Fareed, ``\href{https://doi.org/10.3389/fenvs.2021.814666}{Exchange rate dynamics, energy consumption, and sustainable environment in Pakistan: new evidence from nonlinear ARDL cointegration}'', \textit{Front. Environ. Sci.} \textbf{9} (2022), 1--11.
		
		
		\bibitem{seforall2019}
		Sustainable Energy for All, 
		``\href{https://www.seforall.org/publications/energizing-finance-taking-the-pulse-2019}{Taking the pulse of energy access in Madagascar}'', in \textit{Energizing Finance: Taking the Pulse 2019 - Madagascar, the Philippines and Uganda}, Sustainable Energy for All, 2019.
		
		
		\bibitem{SOLIDIS2017}
		SOLIDIS, ``\href{https://www.solidis.org/wp-content/uploads/2021/07/NL-16-1.pdf}{Tourisme durable et énergie renouvelable: solidis a contribué à la journée thématique}'', \textit{Lettre d'information} \textbf{16} (2017), 2--3.
		
		
		\bibitem{Sunetal2023}
		Z.\ Sun \& X.\ Zhang \& Y.\ Gao, ``\href{https://doi.org/10.3390/ijerph20043124}{The impact of financial development on renewable energy consumption: A multidimensional analysis based on global panel data}'', \textit{International journal of environmental research and public health} \textbf{20} (2023), 1--20. 
		
		
		\bibitem{SurroopRaghoo2018}
		D.\ Surroop \& P.\ Raghoo, ``\href{https://doi.org/10.1016/j.rser.2018.02.024}{Renewable energy to improve energy situation in African island states}'',
		\textit{Renewable and Sustainable Energy Reviews} \textbf{88} (2018), 176--183.
		
		
		\bibitem{SUNREF}
		Sustainable Use of Natural Resources and Energy Finance (SUNREF), ``Sunref Madagascar: SUNREF Madagascar est un trait d’union entre tous les acteurs''. Retrieved October 15, 2023, from \url{https://sunref.solidis.org/sunref-madagascar/}
		
		
		\bibitem{Tibshirani1996}
		R.\ Tibshirani, ``\href{https://doi.org/10.1111/j.2517-6161.1996.tb02080.x}{Regression shrinkage and selection via the lasso}'', \textit{Journal of the Royal Statistical Society. Series B} \textbf{58} (1996), 267--288.
		
		
		\bibitem{Tossou2021}
		E.\ Tossou, ``Rio Tinto va consacrer 7,5 milliards \$ jusqu’en 2030, à la réduction de son empreinte carbone'', \textit{Agence Ecofin}. Retrieved October 1, 2023, from \url{https://www.agenceecofin.com/mines/2210-92579-rio-tinto-va-consacrer-7-5-milliards-jusqu-en-2030-a-la-reduction-de-son-empreinte-carbone}
		
		
		\bibitem{WB_ACCESS}
		World Bank, ``Access to electricity, urban (\% of urban population) - Madagascar''. Retrieved August 20, 2023, from \url{https://data.worldbank.org/indicator/EG.ELC.ACCS.UR.ZS?locations=MG}
		
		
		\bibitem{WDI_REC}
		World Bank, ``Renewable energy consumption (\% of total final energy consumption) - Madagascar''. Retrieved August 20, 2023, from \url{https://data.worldbank.org/indicator/EG.FEC.RNEW.ZS?locations=MG}
		
		
		\bibitem{WDI_URB}
		World Bank, ``Urban population - Madagascar''. Retrieved August 20, 2023, from \url{https://data.worldbank.org/indicator/SP.URB.TOTL?locations=MG}
		
		
		\bibitem{Zhang2016}
		Z.\ Zhang, ``\href{https://doi.org/10.21037/atm.2016.03.35}{Variable selection with stepwise and best subset approaches}'', \textit{Annals of translational medicine} \textbf{4} (2016), 1--6. 
		
	\end{thebibliography}
\end{document}